\documentclass[%
 aip,
 jmp,%
 amsmath,amssymb,
 reprint,%
]{revtex4-1}

\usepackage{graphicx}
\usepackage{dcolumn}
\usepackage{bm}
\usepackage[cal=boondox]{mathalfa}

\DeclareMathAlphabet\mathbfcal{OMS}{cmsy}{b}{n}

\begin{document}

\preprint{AIP/123-QED}

\title[Nonlinear Waves and Coherent Structures in ...]{Nonlinear Waves and Coherent Structures in Quasi-neutral Plasmas Excited by External Electromagnetic Radiation}

\thanks{This article is dedicated to the bright memory of Ronald C. Davidson, a remarkable mentor and friend.}

\author{Stephan I. Tzenov} 

\email{stephan.tzenov@eli-np.ro} 

\affiliation{Extreme Light Infrastructure - Nuclear Physics, 077125 Magurele, Bucharest Ilfov County, Romania}



\date{\today}

\begin{abstract}
Starting from the Vlasov-Maxwell equations describing the dynamics of various species in a quasi-neutral plasma, an exact relativistic hydrodynamic closure for a special type of water-bag distributions satisfying the Vlasov equation has been derived. It has been shown that the set of equations for the macroscopic hydrodynamic variables coupled to the wave equations for the self-consistent electromagnetic field is fully equivalent to the Vlasov-Maxwell system. 

Based on the method of multiple scales, a system comprising a vector nonlinear Schrodinger equation for the transverse envelopes of the self-consistent plasma wakefield, coupled to a scalar nonlinear Schrodinger equation for the electron current velocity envelope, has been derived. 

Using the method of formal series of Dubois-Violette, a traveling wave solution of the derived set of coupled nonlinear Schrodinger equations in the case of circular wave polarization has been obtained. This solution is represented as a ratio of two formal Volterra series. The terms of these series can be calculated explicitly to every desired order. 
\end{abstract}

\maketitle

\section{\label{sec:intro}Introduction} 

Plasma and plasma-like media represent a graceful ground for innovative concepts in modern accelerator and beam physics \cite{Dawson,TajDaw}, because it can sustain high electric fields, thus ensuring very efficient acceleration of charged particles on relatively short distances. Theoretical and experimental studies in this area remain one of the most important research and development activities in both laboratory space and astrophysical plasmas. 

Basically, three mechanisms to generate large amplitude electron
plasma waves are presently put into practice, namely the beat wave mechanism \cite{Noble}, the laser-driven wakefield and excitation of plasma wave structures by particle beams propagating in the plasma medium. Recently, increased interest has also attracted the so-called photon acceleration \cite{Mendonca}. 

Plasma-based accelerators have been proposed by the late John Dawson and his coworkers almost 30 years ago. In recent years this area enjoys a very high popularity among the beam and plasma physics community, and extensive investigations are being performed worldwide with a great deal of success. The near future will answer the question whether such machines will become a serious competitor and eventually displace the conventional "mastodon" family. Experimental results obtained by using powerful short-pulsed lasers \cite{Mangles,Geddes,Faure,Leemans,Kneip,WLeemans}, or a relativistic charged particle beam \cite{Chen,Caldwell} as an external plasma excitation source have demonstrated the production of mono-energetic electron beams with good emittance. These impressive results express significant hope for the future and are very promising for the next generation high energy plasma accelerators, thus providing much smaller "table-top" ion and electron accelerator facilities.

The underlying physical principles relevant to laser-plasma interactions and the present state of the art of plasma based charged particle accelerators that are of significant importance in high energy physics and medicine have been described in a number of books, reviews and articles \cite{Macchi,Eliezer,Mulser,Shukla,Joshi,Blue,Mora,Bingham}. The purpose of the present paper is however rather different. In what follows, we shall discuss the formation and the evolution of nonlinear waves and coherent patterns in plasmas under the action of external electromagnetic radiation -- a topic not so widely touched upon in the literature. 

The paper is organized as follows. In Section \ref{sec:basic}, we review the physical principles and the underlying equations on which the subsequent exposition is based on. Following Ref. \citenum{FEL}, we reduce in Section \ref{sec:waterbag} the Vlasov-Maxwell system to an {\it exact closure} of relativistic warm fluid dynamic equations for the plasma species, which are coupled to the wave equations for the radiation field. The fact that this reduction leads to an exact hydrodynamic closure, makes it extremely valuable simplification for the description of the underlying processes governed by the Vlasov-Maxwell system. An interesting feature of the hydrodynamic system thus derived is the fact that the ponderomotive potential together with the pressure law enter the fluid dynamic picture in the form of an effective enthalpy. Using the method of multiple scales, we perform in Section \ref{sec:nonlinwaves} a reduction of the hydrodynamic and wave equations. As a result, we obtain a vector nonlinear Schrodinger equation describing the evolution of the slowly varying amplitude of the transverse electromagnetic plasma wakefield, coupled to a scalar nonlinear Schrodinger equation for the amplitude of the electron current velocity. An approximate traveling wave solution of the coupled nonlinear Schrodinger equations for the case of circular wave polarization has been found in Section \ref{sec:solcnse}. Finally, in Section \ref{sec:conclude}, we draw some conclusions. 

\section{\label{sec:basic}Theoretical Model and Basic Equations}

We start with the description of a plasma comprised of electrons and ions in an external electromagnetic field depending on the coordinates ${\bf X}$ and time $T$, which is represented by the electromagnetic vector potential ${\widetilde{\bf A}}_e = {\bf e}_x {\widetilde{A}}_{ex} + {\bf e}_y {\widetilde{A}}_{ey}$, where ${\bf e}_x = {\left( 1, 0, 0 \right)}$ and ${\bf e}_y = {\left( 0, 1, 0 \right)}$ are the unit vectors in the $x$ and $y$ direction, respectively. The Hamiltonian governing the dynamics of the different species (electrons and ions) labelled by the subscript $a$ can be written as
\begin{equation}
{\cal H}_a = c {\sqrt{m_a^2 c^2 + {\left( {\bf P} - q_a {\widetilde{\bf A}} \right)}^2}} + q_a {\widetilde{\Phi}}, \label{HamilA}
\end{equation}
where $m_a$ and $q_a$ are the rest mass and the charge of a particle of species $a$, respectively, $c$ is the speed of light in vacuum, ${\bf P}$ is the particle's canonical momentum, and ${\widetilde{\bf A}}_w = {\widetilde{\bf A}} - {\widetilde{\bf A}}_e$ and ${\widetilde{\Phi}}$ are the electromagnetic potentials of the self-fields produced by the plasma particles. 

It is convenient to introduce dimensionless variables according to the relations 
\begin{equation}
t = \omega_e T, \qquad {\bf x} = {\frac {\omega_e} {c}} {\bf X}, \qquad {\bf v} = {\frac {\bf V} {c}}, \label{Variables1}
\end{equation}
\begin{equation} 
{\bf p} = {\frac {\bf P} {m_e c}}, \qquad {\bf A} = {\frac {e {\widetilde{\bf A}}} {m_e c}}, \qquad \Phi = {\frac {e {\widetilde{\Phi}}} {m_e c^2}}, \label{Variables2}
\end{equation}
where 
\begin{equation} 
\omega_e^2 = {\frac {e^2 n_{e0}} {\epsilon_0 m_e}}, \label{PlasFreq}
\end{equation}
is the electron plasma frequency and 
\begin{equation} 
\mu_a = {\frac {m_a} {m_e}}, \label{MassRat}
\end{equation}
is the mass aspect ratio with respect to the electron mass. The new scaled Hamiltonian reads as 
\begin{equation} 
H_a = \mu_a {\sqrt{1 + {\frac {1} {\mu_a^2}} {\left( {\bf p} - Z_a {\bf A} \right)}^2}} + Z_a \Phi, \label{ScaledHam}
\end{equation}
where $Z_a$ is the charge state of the species $a$ ${\left( q_a = e Z_a \right)}$. 

In what follows we assume that spatial variations are one-dimensional in nature, so that the partial derivatives $\partial_x = \partial / \partial x = \partial_y = \partial / \partial y = 0$, while $\partial_s = \partial / \partial s$ is generally nonzero. In other words, the physically relevant phenomena occur in the longitudinal direction $s$ only. 

Next, we perform a canonical transformation aimed at eliminating the longitudinal component of the vector potential $A_s$
\begin{equation} 
F_2^{(a)} {\left( {\bf x}, {\widetilde{\bf p}}; t \right)} = x {\widetilde{p}}_x + y {\widetilde{p}}_y + s {\widetilde{p}}_s + Z_a \int {\rm d} s A_s {\left( s; t \right)}. \label{CanonTrans}
\end{equation}
Dropping the tilde in what follows, the transformed Hamiltonian can be cast in the form 
\begin{eqnarray} 
H_a = \mu_a && {\sqrt{1 + {\frac {p_s^2} {\mu_a^2}} + {\frac {1} {\mu_a^2}} {\left( {\bf p}_{\perp} - Z_a {\bf A}_{\perp} \right)}^2}} \nonumber 
\\ 
&& + Z_a {\left[ \Phi + \partial_t \int {\rm d} s A_s {\left( s; t \right)} \right]}. \label{TransHam}
\end{eqnarray}
Here the subscript "$\perp$" corresponds to the transverse $x$ and $y$ components of the canonical coordinates and fields.

Recalling our initial assumption of independence on the transverse coordinates, we can write the Hamilton's equations of motion as 
\begin{equation} 
{\frac {{\rm d} {\bf x}_{\perp}} {{\rm d} t}} = {\frac {{\bf p}_{\perp} - Z_a {\bf A}_{\perp}} {\mu_a \gamma_a}}, \qquad \qquad {\frac {{\rm d} {\bf p}_{\perp}} {{\rm d} t}} = 0, \label{HamiltEqp}
\end{equation}
\begin{equation} 
{\frac {{\rm d} s} {{\rm d} t}} = {\frac {p_s} {\mu_a \gamma_a}}, \qquad \qquad {\frac {{\rm d} p_s} {{\rm d} t}} = - \mu_a {\frac {\partial \gamma_a} {\partial s}} + Z_a {\cal F}, \label{HamiltEqs}
\end{equation}
where 
\begin{equation} 
\gamma_a = {\sqrt{1 + {\frac {p_s^2} {\mu_a^2}} + {\frac {1} {\mu_a^2}} {\left( {\bf p}_{\perp} - Z_a {\bf A}_{\perp} \right)}^2}}, \label{Gammasa}
\end{equation}
and 
\begin{equation} 
{\cal F} = - \partial_s \Phi - \partial_t A_s, \label{ElecForce}
\end{equation}
is the electric force acting on a unit-charge particle in the longitudinal direction. From the second set of Hamilton's equations (\ref{HamiltEqp}) it follows that there are two exact single-particle invariants in the combined external and self-field configuration. These are the canonical momenta ${\bf p}_{\perp}$, transverse to the beam propagation direction.

The nonlinear Vlasov equation for the distribution function $f_a {\left( {\bf x}, {\bf p}; τ \right)}$ of particles of species $a$ can be written as 
\begin{eqnarray} 
\partial_t f_a + && {\frac {{\bf p}_{\perp} - Z_a {\bf A}_{\perp}} {\mu_a \gamma_a}} \cdot {\boldsymbol{\nabla}}_{\perp} f_a + {\frac {p_s} {\mu_a \gamma_a}} \partial_s f_a \nonumber 
\\ 
&& + {\left( Z_a {\cal F} - \mu_a \partial_s \gamma_a \right)} \partial_{p_s} f_a = 0, \label{Vlasov}
\end{eqnarray}
where ${\boldsymbol{\nabla}}_{\perp} = {\left( \partial_x, \partial_y \right)}$ denotes the transverse components of the well-known gradient operator. Using the fact that the transverse canonical momenta are integrals of motion, it is a matter of straightforward verification that it possesses a solution of the form 
\begin{equation} 
f_a {\left( {\bf x}, {\bf p}; τ \right)} = \delta {\left( p_x \right)} \delta {\left( p_y \right)} F_a {\left( s, p_s; t \right)}. \label{VlasovSol}
\end{equation}
The evolution of the yet unknown function $F_a {\left( s, p_s; t \right)}$ depending on the longitudinal canonical coordinates and the time is governed by the one-dimensional Vlasov equation
\begin{equation} 
\partial_t F_a + {\frac {p_s} {\mu_a \gamma_a}} \partial_s F_a + {\left( Z_a {\cal F} - \mu_a \partial_s \gamma_a \right)} \partial_{p_s} F_a = 0, \label{VlasovOneDF}
\end{equation}
where the Lorentz factor (dimensionless kinetic energy of a unit mass particle with a zero transverse canonical momentum) $\gamma_a$ is defined now according to the relation 
\begin{equation} 
\gamma_a {\left( s, p_s; t \right)} = {\sqrt{1 + {\frac {p_s^2} {\mu_a^2}} + {\frac {Z_a^2} {\mu_a^2}} A^2 {\left( s; t \right)}}}, \qquad A^2 = A_x^2 + A_y^2. \label{Gammakin}
\end{equation}
Expression (\ref{VlasovSol}) implies that the plasma species are "cold" in the transverse direction, since their transverse temperature has been neglected. 

\section{\label{sec:waterbag}Longitudinal Water Bag Model}

In this Section we shall follow closely Ref. \citenum{FEL} by selecting a class of exact solutions to the Vlasov equation (\ref{VlasovOneDF}) given by the expression 
\begin{eqnarray} 
F_a {\left( s, p_s; t \right)} = {\cal C}_a && {\left\{ \Theta {\left[ p_s - p_a^{(-)} {\left( s, t \right)} \right]} \right.} \nonumber
\\ 
&& {\left. - \Theta {\left[ p_s - p_a^{(+)} {\left( s, t \right)} \right]} \right\}}, \label{Waterbag}
\end{eqnarray}
where $\Theta {\left( z \right)}$ is the well known Heaviside step function. Further details about the basic properties and the underlying physical meaning of the class of uniform-phase-density (water bag) distribution functions, which solve exactly the one-degree-of-freedom Vlasov equation can be found in Refs. \citenum{TzenovBOOK} and \citenum{PRSTAB}.

The equations governing the dynamics of the water bag boundary curves can be written as 
\begin{equation} 
\partial_t {\left( p_a^{(+)} - p_a^{(-)} \right)} + \mu_a \partial_s {\left( \gamma_a^{(+)} - \gamma_a^{(-)}  \right)} = 0, \label{WBContin}
\end{equation}
\begin{eqnarray} 
{\frac {1} {2}} \partial_t {\left( p_a^{(+) {\bf 2}} - p_a^{(-) {\bf 2}} \right)} + && \mu_a {\left( p_a^{(+)} \partial_s \gamma_a^{(+)} - p_a^{(-)} \partial_s \gamma_a^{(-)}  \right)} \nonumber
\\
&& = Z_a {\cal F} {\left( p_a^{(+)} - p_a^{(-)} \right)}, \label{WBMomBal}
\end{eqnarray}
where in accordance with Eq. (\ref{Gammakin}), we have defined
\begin{equation} 
\gamma_a^{(\pm)} {\left( s; t \right)} = {\sqrt{1 + {\frac {1} {\mu_a^2}} {\left[ p_a^{(\pm) {\bf 2}} {\left( s; t \right)} + Z_a^2 A^2 {\left( s; t \right)} \right]}}}. \label{GammaWB}
\end{equation}
In analogy with what has been done in Ref. \citenum{FEL}, it is possible to introduce hydrodynamic variables $n_a$, $V_a$ and $\Gamma_a$ for each plasma species. Omitting details, which can be reproduced in a straightforward manner, we state here the final result. The important quantity $\Gamma_a$ can be written as 
\begin{equation} 
\Gamma_a = {\sqrt{\frac {1 + {\dfrac {Z_a^2} {\mu_a^2}} A^2} {{\left( 1 - V_a^2 \right)} {\left( 1 - 2 v_{aT}^2 n_a^2 \right)}}}}, \label{GammaSpeca}
\end{equation}
where 
\begin{equation} 
v_{aT}^2 = {\frac {1} {8 {\cal C}_a^2}}, \label{ThermSpeed}
\end{equation}
is the thermal speed squared of the plasma species of the type $a$. 

The completion of the macroscopic fluid description can be performed in a similar to Ref. \citenum{FEL} manner by expressing the source terms entering the corresponding wave equations for the electromagnetic potentials as functions of $n_a$, $V_a$ and $\Gamma_a$. Using the Lorentz gauge
\begin{equation} 
\partial_t \Phi + {\boldsymbol{\nabla}} \cdot {\bf A} = 0, \label{LorentzGauge}
\end{equation}
in the dimensionless variables introduced in the preceding Section, we can write the constitutive macroscopic hydrodynamic equation for each plasma species $a$ coupled with the wave equations for the self-consistent fields as 
\begin{equation} 
\partial_t {\left( n_a \Gamma_a \right)} + \partial_s {\left( n_a \Gamma_a V_a  \right)} = 0, \label{ContinuitySpA}
\end{equation}
\begin{equation} 
\partial_t {\left( V_a \Gamma_a \right)} + \partial_s \Gamma_a = {\cal F}_a = - {\frac {Z_a} {\mu_a}} {\left( \partial_s \Phi + \partial_t A_s \right)}, \label{MomBalSpA}
\end{equation}
\begin{equation} 
{\boldsymbol{\Box}} \Phi = - {\frac {1} {n_{e0}}} \sum \limits_a Z_a n_a \Gamma_a, \label{WaveEqScalar}
\end{equation}
\begin{equation} 
{\boldsymbol{\Box}} A_s = - {\frac {1} {n_{e0}}} \sum \limits_a Z_a n_a \Gamma_a V_a, \label{WaveEqVecLong}
\end{equation}
\begin{equation} 
{\boldsymbol{\Box}} {\bf A}_{\perp} = {\frac {{\bf A}_{\perp}} {n_{e0}}} \sum \limits_a Z_a^2 n_a {\left( 1 + {\frac {2} {3}} v_{aT}^2 n_a^2 \right)} + \Box {\bf A}_e. \label{WaveEqVecTrans}
\end{equation}
Here ${\boldsymbol{\Box}} = \partial_s^2 - \partial_t^2$ is the well known d'Alembert operator. 

To summarize, for the case of constant phase-space density distribution in Eq. (\ref{Waterbag}), the macroscopic fluid description provided by Eqs. (\ref{ContinuitySpA}) -- (\ref{WaveEqVecTrans}), is fully equivalent to the nonlinear Vlasov-Maxwell equations (\ref{VlasovOneDF}) and the corresponding wave equations for the self fields. This remarkable simplification, i.e., exact closure of the hydrodynamic equations with the first two velocity moments is a consequence of the fact that the heat flow can be shown to be zero \cite{FEL} for the class of relativistic water-bag distribution functions. It is worth be pointed out that Eqs. (\ref{ContinuitySpA}) -- (\ref{WaveEqVecTrans}), are readily amenable to numerical solution, and can also be investigated analytically, which will be done in the subsequent Sections. 

\section{\label{sec:nonlinwaves}Nonlinear Waves and Coherent Structures}

\subsection{\label{subsec:basequations}Derivation of the Basic Equations}

Before we proceed, let us mention an important property of our basic system of hydrodynamic and wave equations (\ref{ContinuitySpA}) -- (\ref{WaveEqVecTrans}). Since the external pumping electromagnetic field satisfies the homogeneous Maxwell equations ${\left( {\boldsymbol{\Box}} {\bf A}_e = 0 \right)}$, the system (\ref{ContinuitySpA}) -- (\ref{WaveEqVecTrans}) possesses an obvious stationary solution 
\begin{equation} 
n_a = n_{as} = {\rm const}, \qquad \Gamma_a = \Gamma_{as} = {\rm const}, \qquad V_{as} = 0, \label{StatSol1}
\end{equation}
\begin{equation} 
{\cal F}_{as} = 0, \qquad {\bf A}_s = 0, \qquad \Phi_s = 0, \label{StatSol2}
\end{equation}
provided the quasi-neutrality condition 
\begin{equation} 
\sum \limits_a Z_a n_{as} \Gamma_{as} = 0, \label{Quasineutral}
\end{equation}
holds. Moreover, ions comprise a heavy plasma background, so that their effect on the formation and the dynamics of the plasma wakefield, triggered by the external pumping electromagnetic field can be neglected. Thus, taking into account the contribution from electrons only, Eqs. (\ref{ContinuitySpA}) -- (\ref{WaveEqVecTrans}) can be rewritten as
\begin{equation} 
\partial_t {\left( n \Gamma \right)} + \partial_s {\left( n \Gamma V  \right)} = 0, \label{Continuity}
\end{equation}
\begin{equation} 
\partial_t {\left( \Gamma V \right)} + \partial_s \Gamma = {\cal F} = \partial_s \Phi + \partial_t A_s, \label{MomBalance}
\end{equation}
\begin{equation} 
{\boldsymbol{\Box}} \Phi = n \Gamma, \label{WaveEqScal}
\end{equation}
\begin{equation} 
{\boldsymbol{\Box}} A_s = n \Gamma V, \label{WaveEqVecs}
\end{equation}
\begin{equation} 
{\boldsymbol{\Box}} {\bf A}_{\perp} = n {\left( 1 + {\frac {2} {3}} v_T^2 n^2 \right)} {\bf A}_{\perp}. \label{WaveEqVecPerp}
\end{equation}
Here the scaling $n = n_e / n_{e0}$ and $v_T = n_{e0} v_{eT}$ has been used. 

It is possible to eliminate the two wave equations (\ref{WaveEqScal}) and (\ref{WaveEqVecs}) for the scalar potential $\Phi$ and for the longitudinal component $A_s$ of the vector potential. To do so we apply the d'Alembert operator to both sides of the equation ${\cal F} = \partial_s \Phi + \partial_t A_s$ and obtain 
\begin{equation} 
{\boldsymbol{\Box}} {\cal F} = \partial_s {\left( n \Gamma \right)} + \partial_t {\left( n \Gamma V \right)}. \label{ForceEquat}
\end{equation}
Further, we differentiate the momentum balance equation (\ref{MomBalance}) with respect to the time $t$ and then apply the d'Alembert operator to both sides. Taking into account Eq. (\ref{ForceEquat}) and the continuity equation (\ref{Continuity}), we finally arrive at 
\begin{equation} 
{\boldsymbol{\Box}} {\left[ \partial_t^2 {\left( \Gamma V \right)} + \partial_t \partial_s \Gamma \right]} = - \Box {\left( n \Gamma V \right)}. \label{MomBalModif}
\end{equation}
This last equation supplemented by the continuity equation (\ref{Continuity}) and the wave equation (\ref{WaveEqVecPerp}) for the transverse components of the vector potential constitutes the starting point for our subsequent analysis. 

\subsection{\label{subsec:multiple}Reduction by the Method of Multiple Scales}

For the sake of convenience, let us introduce the following notations  
\begin{equation} 
\Gamma = \gamma G_A, \label{Notation1}
\end{equation}
where
\begin{equation} 
\gamma = {\frac {1} {\sqrt{1 - V^2}}}, \qquad \quad G_A = {\sqrt{\frac {1 + A^2} {1 - 2 v_T^2 n^2}}}, \label{Notation1a}
\end{equation}
and 
\begin{equation} 
A^2 = A_x^2 + A_y^2. \label{Notation2}
\end{equation}

Following the standard procedure of the multiple scales reduction method \cite{TzenovBOOK,Nayfeh,Kevorkian} applied to the system of equations (\ref{Continuity}), (\ref{WaveEqVecPerp}) and (\ref{MomBalModif}), we represent the line density $n$, the current velocity $V$ and the transverse vector potential ${\bf A}_{\perp}$ as a perturbation expansion according to the expressions  
\begin{equation} 
n = 1 + \sum \limits_{k=1}^{\infty} \epsilon^k n_k, \quad V = \sum \limits_{k=1}^{\infty} \epsilon^k v_k, \quad {\bf A}_{\perp} = \sum \limits_{k=1}^{\infty} \epsilon^k {\bf A}_k. \label{DenseExpand}
\end{equation}
Here $\epsilon$ is a formal small parameter, which will be set equal to one at the end of all calculations. In addition, the differential operators with respect to the time $t$ and to the longitudinal spatial variable $s$ are also expanded in the small parameter $\epsilon$ as follows
\begin{equation}
\partial_t = \sum \limits_{n=0}^{\infty} \epsilon^n \partial_{t_n}, \qquad \qquad \partial_s = \sum \limits_{n=0}^{\infty} \epsilon^n \partial_{s_n}, \label{PerturbOper}
\end{equation}
where 
\begin{equation}
t_n = \epsilon^n t, \qquad \qquad s_n = \epsilon^n s. \label{PerturbVar}
\end{equation}
The next step consists in expanding the system of hydrodynamic and field equations (\ref{Continuity}), (\ref{WaveEqVecPerp}) and (\ref{MomBalModif}) in the formal small parameter $\epsilon$. Their perturbation solution will be obtained order by order together with performing a procedure of elimination of secular terms (starting with second order), which will yield the sought for amplitude equations for the slowly varying envelopes. 

\subsubsection{\label{subsubsec:firstorder}First Order} 

The linearized Eqs. (\ref{Continuity}), (\ref{WaveEqVecPerp}) and (\ref{MomBalModif}) can be written as 
\begin{equation}
G_0^2 \partial_t n_1 + \partial_s v_1 = 0, \label{LinContin}
\end{equation}
\begin{equation}
\partial_t^2 v_1 + 2 v_T^2 G_0^2 \partial_t \partial_s n_1 = - v_1, \label{LinMombal}
\end{equation}
\begin{equation}
{\boldsymbol{\Box}} {\bf A}_1 = \lambda {\bf A}_1, \label{LinWave}
\end{equation}
where 
\begin{equation}
G_0 = {\frac {1} {\sqrt{1 - 2 v_T^2}}}, \qquad \qquad \lambda = 1 + {\frac {2} {3}} v_T^2. \label{Params}
\end{equation}
The solutions to the equations 
\begin{equation}
{\left( \partial_t^2 - 2 v_T^2 \partial_s^2 + 1 \right)} v_1 = 0, \quad \quad {\left( {\boldsymbol{\Box}} - \lambda \right)} {\bf A}_1 = 0, \label{LinearEquat}
\end{equation}
for the first order current velocity $v_1$ and for the first order transverse vector potential ${\bf A}_1$ are 
\begin{equation}
v_1 = {\cal B} {\rm e}^{i \varphi} + {\cal B}^{\ast} {\rm e}^{- i \varphi}, \quad \quad {\bf A}_1 = {\mathbfcal A} {\rm e}^{i \psi} + {\mathbfcal A}^{\ast} {\rm e}^{- i \psi}. \label{FirstOrdSol}
\end{equation}
The wave phases of the two waves propagating in the longitudinal direction can be expressed as 
\begin{equation}
\varphi = k s - \Omega t, \qquad \qquad \psi = k s - \omega t, \label{WavePhase}
\end{equation}
where the wave frequencies are given by 
\begin{equation}
\Omega = {\sqrt{1 + 2 k^2 v_T^2}} \qquad \qquad \omega = {\sqrt{k^2 + \lambda}}. \label{WaveFreq}
\end{equation}
It is important to emphasize that the wave amplitudes ${\cal B}$ and ${\mathbfcal A} = {\cal A}_x {\bf e}_x + {\cal A}_y {\bf e}_y$ are constants with respect to the fast scales $t$ and $s$, but they can depend on the slower scales $t_1, t_2, \dots$ and $s_1, s_2, \dots$ in general. Solving Eq. (\ref{LinContin}), we obtain 
\begin{equation}
n_1 = {\frac {k} {G_0^2 \Omega}} {\left( {\cal B} {\rm e}^{i \varphi} + {\cal B}^{\ast} {\rm e}^{- i \varphi} \right)}. \label{FirstOrdDensity}
\end{equation}

\subsubsection{\label{subsubsec:secorder}Second Order} 

The explicit form of the second order equations is given in Appendix \ref{sec:appsecthird}. Close inspection of Eqs. (\ref{SecOrdWavePerp}) and (\ref{SecOrdMomBalan}) indicates that the second terms on their left-hand-sides would give rise to unwanted secular terms, which linearly grow in time. In order to avoid such nonphysical artifact due to the naive perturbation solution, we can use the additional degree of freedom provided by the method of multiple scales and require that the above-mentioned terms vanish identically. This yields the first-order amplitude equations (or usually called the solvability conditions), which can be written as 
\begin{equation}
{\left( \partial_{t_1} + v_{\Omega} \partial_{s_1} \right)} {\cal B} = 0, \qquad \quad {\left( \partial_{t_1} + v_{\omega} \partial_{s_1} \right)} {\mathbfcal A} = 0, \label{FirstOrdSolCond}
\end{equation}
where 
\begin{equation}
v_{\Omega} = {\frac {{\rm d} \Omega} {{\rm d} k}} = {\frac {2 k v_T^2} {\Omega}}, \qquad \quad v_{\omega} = {\frac {{\rm d} \omega} {{\rm d} k}} = {\frac {k} {\omega}}, \label{GroupVeloc}
\end{equation}
are the group velocities of the two waves \cite{TzenovTUBES}. 

Thus, the general solution to our second-order perturbation equations acquires the form 
\begin{equation}
{\bf A}_2 = \alpha_2 {\cal B} {\mathbfcal A} {\rm e}^{i {\left( \varphi + \psi  \right)}} + \beta_2 {\cal B} {\mathbfcal A}^{\ast} {\rm e}^{i {\left( \varphi - \psi  \right)}} + c.c., \label{SeOrdSolPot}
\end{equation}
\begin{equation}
v_2 = \gamma_2 {\cal A}^2 {\rm e}^{2 i \psi} + {\frac {\Delta_2} {3}} {\cal B}^2 {\rm e}^{2 i \varphi} + c.c., \label{SeOrdSolVel}
\end{equation}
where 
\begin{equation}
\alpha_2 = {\frac {k {\left( 1 + 2 v_T^2 \right)}} {G_0^2 \Omega {\left( \Omega^2 + 2 \omega \Omega - 3 k^2 \right)}}} \label{Coefficient1}
\end{equation}
\begin{equation}
\beta_2 = {\frac {k {\left( 1 + 2 v_T^2 \right)}} {G_0^2 \Omega {\left( \Omega^2 - 2 \omega \Omega + k^2 \right)}}}, \label{Coefficient2}
\end{equation}
\begin{equation}
\gamma_2 = {\frac {2 \omega k} {4 k^2 + {\left( 4 \lambda - 1 \right)} G_0^2}}, \label{Coefficient3}
\end{equation}
\begin{equation}
\Delta_2 = {\frac {2 k \Omega} {G_0^2}} {\left( 1 + {\frac {2 k^2 v_T^2} {\Omega^2}} \right)} + {\frac {k} {\Omega}} {\left[ 1 - 8 v_T^2 {\left( \Omega^2 - k^2 \right)} \right]}, \label{Coefficient4}
\end{equation}
and "$c.c.$" implies complex conjugation. For the second order number density, we obtain 
\begin{eqnarray}
n_2 = - {\frac {i} {G_0^2 \Omega^2}} && {\left( k \partial_{t_1} + \Omega \partial_{s_1} \right)} {\cal B} {\rm e}^{i \varphi} + {\frac {\lambda_2} {2 G_0^2 \Omega}} {\cal B}^2 {\rm e}^{2 i \varphi} \nonumber
\\ 
&& + {\frac {2 k \gamma_2 - \omega} {2 G_0^2 \omega}} {\cal A}^2 {\rm e}^{2 i \psi} + c.c., \label{SeOrdSolDens}
\end{eqnarray}
where 
\begin{equation}
\lambda_2 = {\frac {2 k \Delta_2} {3}} + {\frac {2 k^2 {\left( 1 - 3 v_T^2 \right)}} {\Omega}} - \Omega. \label{Coefficient5}
\end{equation}

\subsubsection{\label{subsubsec:thiorder}Third Order - Derivation of the Amplitude Equations} 

In third order, we retain only secular (resonant) terms, which follow the pattern of the two basic plasma waves (with phases ${\rm e}^{i \varphi}$ or  ${\rm e}^{i \psi}$, respectively). The rest contribute to the regular solution of the third order perturbation equations, involving higher harmonics and combinations of higher order of the basic plasma modes. The condition for elimination of the above-mentioned secular contribution from the general perturbation solution of our initial system (\ref{Continuity}), (\ref{WaveEqVecPerp}) and (\ref{MomBalModif}), yields the sought for amplitude equations. Omitting straightforwardly reproducible calculation's details, we write down the final result 
\begin{equation}
i \partial_{t_2} {\mathbfcal A} + i v_{\omega} \partial_{s_2} {\mathbfcal A} = - {\frac {1} {2}} {\frac {{\rm d} v_{\omega}} {{\rm d} k}} \partial_{s_1}^2 {\mathbfcal A} + \Gamma_{aa} {\cal A}^2 {\mathbfcal A}^{\ast} + \Gamma_{ab} {\left| {\cal B} \right|}^2 {\mathbfcal A}, \label{NonLinSchrodA}
\end{equation}
\begin{equation}
i \partial_{t_2} {\cal B} + i v_{\Omega} \partial_{s_2} {\cal B} = - {\frac {1} {2}} {\frac {{\rm d} v_{\Omega}} {{\rm d} k}} \partial_{s_1}^2 {\cal B} + \Gamma_{ba} {\left| {\cal A} \right|}^2 {\cal B} + \Gamma_{bb} {\left| {\cal B} \right|}^2 {\cal B}. \label{NonLinSchrodB}
\end{equation}
Here, the following notations
\begin{equation}
{\frac {{\rm d} v_{\omega}} {{\rm d} k}} = {\frac {1} {\omega}} {\left( 1 - v_{\omega}^2 \right)}, \label{Notatio1}
\end{equation}
\begin{equation}
\Gamma_{aa} = {\frac {1} {4 \omega^2 G_0^2}} {\left( 1 + 2 v_T^2 \right)} {\left( 2 k \gamma_2 - \omega \right)}, \label{Notatio2}
\end{equation}
\begin{equation}
\Gamma_{ab} = {\frac {1} {2 \omega}} {\left[ {\frac {4 k^2 v_T^2} {G_0^4 \Omega^2}} + {\frac {k {\left( 1 + 2 v_T^2 \right)}} {G_0^2 \Omega}} {\left( \alpha_2 + \beta_2 \right)} \right]}, \label{Notatio3}
\end{equation}
and 
\begin{equation}
{\frac {{\rm d} v_{\Omega}} {{\rm d} k}} = {\frac {1} {\Omega}} {\left( 2 v_T^2 - v_{\Omega}^2 \right)}, \label{Notatio12}
\end{equation}
\begin{equation}
\Gamma_{ba} = {\frac {1} {\Omega}} + {\frac {k} {2 G_0^2}} {\left( \alpha_2 + \beta_2  \right)}, \label{Notatio22}
\end{equation}
\begin{eqnarray}
\Gamma_{bb} = && {\frac {1} {2 \Omega}} {\left[ 3 + {\frac {k \Omega \Delta_2} {3 G_0^2}} + {\frac {1} {\Omega}} {\left( {\frac {\lambda_2} {2}} + {\frac {k \Delta_2} {3}} \right)} + {\frac {6 k^2 v_T^2} {G_0^2}} \right.} \nonumber 
\\ 
&& {\left. + {\frac {v_T^2} {\Omega}} {\left( k^2 - \Omega^2 \right)} {\left( \lambda_2 + {\frac {2} {3}} k \Delta_2 \right)} \right.} \nonumber  
\\ 
&& {\left. + {\frac {12 k^4 v_T^4} {G_0^2 \Omega^2}} + {\frac {v_T^2 k^2 \lambda_2} {G_0^2 \Omega}} \right]}, \label{Notatio32}
\end{eqnarray}
have been used. Moreover, ${\cal A}^2 = {\mathbfcal A} \cdot {\mathbfcal A}$ is complex, while ${\left| {\cal A} \right|}^2 = {\mathbfcal A} \cdot {\mathbfcal A}^{\ast}$ is real. As already mentioned in Section \ref{subsec:multiple}, the formal small parameter $\epsilon$ will be set equal to one, so that $s_1 = s$, $s_2 = s$ and $t_2 = t$.

Equations (\ref{NonLinSchrodA}) and (\ref{NonLinSchrodB}) comprise a system of a coupled nonlinear vector Schrodinger equations for ${\mathbfcal A}$ and a scalar nonlinear Schrodinger equation for ${\cal B}$. They describe the evolution of the slowly varying amplitudes of the generated transverse plasma wakefield and the current velocity of the plasma electrons. 

An important remark is now in order. Without loss of generality, we can express the vertical component ${\cal A}_y$ of the plasma wakefield in terms of the horizontal one ${\cal A}_x$ according to the relation  
\begin{equation}
{\cal A}_y = {\cal A}_x {\cal C}, \label{CalARepres}
\end{equation}
where ${\cal C}$ is some complex valued function. Then, the vector nonlinear Schrodinger equation (\ref{NonLinSchrodA}) splits into two equations as follows 
\begin{eqnarray}
i \partial_t {\cal A}_x && + i v_{\omega} \partial_s {\cal A}_x = - {\frac {v_{\omega}^{\prime}} {2}} \partial_s^2 {\cal A}_x \nonumber 
\\ 
&& + \Gamma_{aa} {\left( 1 + {\cal C}^2 \right)} {\left| {\cal A}_x \right|}^2 {\cal A}_x + \Gamma_{ab} {\left| {\cal B} \right|}^2 {\cal A}_x, \label{NoLinSchrodAx}
\end{eqnarray}
\begin{eqnarray}
i \partial_t {\cal C} + i v_{\omega} \partial_s {\cal C} && = - {\frac {v_{\omega}^{\prime}} {2}} {\left[ \partial_s^2 {\cal C} + 2 {\left( \partial_s \ln {\cal A}_x \right)} \partial_s {\cal C} \right]} \nonumber 
\\ 
&& + \Gamma_{aa} {\left( 1 + {\cal C}^2 \right)} {\left| {\cal A}_x \right|}^2 {\left( {\cal C}^{\ast} - {\cal C} \right)}. \label{NoLinSchrodC}
\end{eqnarray}
For the sake of simplicity, here and in what follows the shorthand notations $v_{\omega}^{\prime} = {\rm d} v_{\omega} / {\rm d} k$ and $v_{\Omega}^{\prime} = {\rm d} v_{\Omega} / {\rm d} k $ have been introduced. Note that the second Eq. (\ref{NoLinSchrodC}) of the above equations possesses a stationary solution ${\cal C} = {\rm const}$, so that the vector Schrodinger equation (\ref{NonLinSchrodA}) degenerates into a single scalar equation (\ref{NoLinSchrodAx}). The two possibilities are: 
\begin{itemize}
 \item ${\cal C} = {\mathcal p}$ is real. The incident ${\mathcal p} = 0$ corresponds to the physically relevant case of linear wave polarization. 
 \item ${\cal C} = \pm i$. This corresponds to circular wave polarization. 
\end{itemize} 

The system of nonlinear Schrodinger equations (\ref{NonLinSchrodA}) and (\ref{NonLinSchrodB}) can be simplified considerably by introducing new independent variables $\xi$ and $\eta$ according to the relations 
\begin{equation}
\xi = {\mathcal a} {\left( s - v_{\omega} t \right)}, \quad \eta = - {\mathcal a} {\left( s - v_{\Omega} t \right)},\quad {\mathcal a} = {\frac {1} {v_{\Omega} - v_{\omega}}}. \label{IndVarxieta}
\end{equation}
Taking into account the expressions for the derivatives 
\begin{equation}
\partial_t = - {\mathcal a} {\left( v_{\omega} \partial_{\xi} - v_{\Omega} \partial_{\eta} \right)}, \qquad \partial_s = {\mathcal a} {\left( \partial_{\xi} - \partial_{\eta} \right)}, \label{DerivNewVar}
\end{equation}
we cast Eqs. (\ref{NonLinSchrodA}) and (\ref{NonLinSchrodB}) into the form 
\begin{equation}
i \partial_{\eta} {\mathbfcal A} = - {\frac {{\mathcal a}^2 v_{\omega}^{\prime}} {2}} {\left( \partial_{\xi} - \partial_{\eta} \right)}^2 {\mathbfcal A} + \Gamma_{aa} {\cal A}^2 {\mathbfcal A}^{\ast} + \Gamma_{ab} {\left| {\cal B} \right|}^2 {\mathbfcal A}, \label{NLinSchrodA}
\end{equation}
\begin{equation}
i \partial_{\xi} {\cal B} = - {\frac {{\mathcal a}^2 v_{\Omega}^{\prime}} {2}} {\left( \partial_{\xi} - \partial_{\eta} \right)}^2 {\cal B} + \Gamma_{ba} {\left| {\cal A} \right|}^2 {\cal B} + \Gamma_{bb} {\left| {\cal B} \right|}^2 {\cal B}. \label{NLinSchrodB}
\end{equation}

Although not yet verified explicitly, the most natural conjecture is that the above system of equations in its generic form is not integrable. There might exist cases for particular values of the $\Gamma$-coupling coefficients depending on the wave number $k$, where integrability can be proved, but these remain to be specifically investigated. Therefore, in order to understand the physical essence underlying these equations, one has to employ adequate methods to seek for approximate solutions (at least), and/or to tackle the above system numerically. A potential candidate for a compassable analytical treatment of the set of nonlinear Schrodinger equations (\ref{NLinSchrodA}) and (\ref{NLinSchrodB}) is the non-conventional Hirota's bilinear method outlined in Appendix \ref{sec:apphirota}. The correct bilinear equations can be subsequently solved perturbatively by following a standard and commonly adopted procedure. 

\section{\label{sec:solcnse}Traveling Wave Solution of the Coupled Nonlinear Schrodinger Equations (\ref{NLinSchrodA}) and (\ref{NLinSchrodB})}

Instead of utilizing the non-conventional Hirota's bilinear method, we shall introduce in what follows a new technique, which yields a similar result. The so-called method of formal series described below is in our opinion more straightforward and simpler to use.  

We shall focus here on the analysis of circularly polarized plasma waves, in which case Eqs. (\ref{NLinSchrodA}) and (\ref{NLinSchrodB}) can be written as 
\begin{equation}
i \partial_{\eta} {\cal A}_x = - {\frac {{\mathcal a}^2 v_{\omega}^{\prime}} {2}} {\left( \partial_{\xi} - \partial_{\eta} \right)}^2 {\cal A}_x + \Gamma_{ab} {\left| {\cal B} \right|}^2 {\cal A}_x, \label{NLinSchrcpA}
\end{equation}
\begin{equation}
i \partial_{\xi} {\cal B} = - {\frac {{\mathcal a}^2 v_{\Omega}^{\prime}} {2}} {\left( \partial_{\xi} - \partial_{\eta} \right)}^2 {\cal B} + 2 \Gamma_{ba} {\left| {\cal A}_x \right|}^2 {\cal B} + \Gamma_{bb} {\left| {\cal B} \right|}^2 {\cal B}. \label{NLinSchrcpB}
\end{equation}
The case of linear wave polarization can be treated in a similar manner. 

Traveling wave solutions are generally sought through the standard ansatz 
\begin{equation}
{\cal A}_x {\left( \xi, \eta \right)} = {\rm e}^{i {\left( \mu \xi + \nu_1 \eta \right)}} {\cal P} {\left( z \right)}, \quad {\cal B} {\left( \xi, \eta \right)} = {\rm e}^{i {\left( \mu \xi + \nu_2 \eta \right)}} {\cal Q} {\left( z \right)}, \label{TravWaveAnsatz}
\end{equation}
where $z = \eta - u \xi$ is a new variable, while ${\cal P}$ and ${\cal Q}$ are yet unknown complex (in general) traveling wave amplitudes. The quantities $\mu$, $\nu_{1,2}$ and the traveling wave velocity $u$ are constants to be determined additionally. Without loss of generality, we can set 
\begin{equation}
u = 0, \quad \qquad \nu_1 = \mu - {\frac {1} {{\mathcal a}^2 v_{\omega}^{\prime}}}, \quad \qquad \nu_2 = \mu. \label{SetConst}
\end{equation}
Thus, Eqs. (\ref{NLinSchrcpA}) and (\ref{NLinSchrcpB}) can be expressed as 
\begin{equation}
{\frac {{\rm d}^2 {\cal P}} {{\rm d} \eta^2}} - \omega_1^2 {\cal P} = {\frac {2 \Gamma_{ab}} {{\mathcal a}^2 v_{\omega}^{\prime}}} {\left| {\cal Q} \right|}^2 {\cal P}, \label{NLinODEP}
\end{equation}
\begin{equation}
{\frac {{\rm d}^2 {\cal Q}} {{\rm d} \eta^2}} - \omega_2^2 {\cal Q} = {\frac {4 \Gamma_{ba}} {{\mathcal a}^2 v_{\Omega}^{\prime}}} {\left| {\cal P} \right|}^2 {\cal Q} + {\frac {2 \Gamma_{bb}} {{\mathcal a}^2 v_{\Omega}^{\prime}}} {\left| {\cal Q} \right|}^2 {\cal Q}, \label{NLinODEQ}
\end{equation}
where 
\begin{equation}
\omega_1 = {\sqrt{{\frac {2 \mu} {{\mathcal a}^2 v_{\omega}^{\prime}}} - {\frac {1} {{\mathcal a}^4 v_{\omega}^{\prime {\bf 2}}}}}}, \qquad \quad \omega_2 = {\sqrt{\frac {2 \mu} {{\mathcal a}^2 v_{\Omega}^{\prime}}}}. \label{FreqOmega12}
\end{equation}
Now, we can consider the traveling wave amplitudes ${\cal P}$ and ${\cal Q}$ real. The system (\ref{NLinODEP}) and (\ref{NLinODEQ}) comprises a set of two nonlinearly coupled Duffing equations, which is known to exhibit chaotic behavior. Note also that for negative values of the free parameter $\mu$, the quantities $\omega_1$ and $\omega_2$ become purely imaginary, and Eqs. (\ref{NLinODEP}) and (\ref{NLinODEQ}) represent a system of coupled Duffing oscillators. 

The system of nonlinearly coupled Duffing equations (\ref{NLinODEP}) and (\ref{NLinODEQ}) can be transformed to equivalent form as a set of coupled nonlinear Volterra integral equations of the second kind 
\begin{equation}
{\bf P} {\left( \eta \right)} - {\mathbfcal G} {\left[ \eta; {\bf P} {\left( \eta \right)} \right]} = {\bf P}_0 {\left( \eta \right)}. \label{VolterraVec}
\end{equation}
Here, the following vector notations 
\begin{equation}
{\bf P} = {\left( {\cal P}, {\cal Q} \right)}, \qquad \quad {\bf P}_0 = {\left( {\cal P}_0, {\cal Q}_0 \right)}, \label{VecNotat}
\end{equation}
have been introduced. The components of the vector functional ${\mathbfcal G} = {\left( {\cal G}_P, {\cal G}_Q \right)}$ are explicitly expressed according to the relations   
\begin{equation}
{\cal G}_P {\left[ \eta; {\bf P} {\left( \eta \right)} \right]} = {\frac {2 p_0} {\omega_1}} \int \limits_{0}^{\eta} {\rm d} \sigma \sinh \omega_1 {\left( \eta - \sigma \right)} {\cal Q}^2 {\left( \sigma \right)} {\cal P} {\left( \sigma \right)}, \label{VecFunctP}
\end{equation}
\begin{eqnarray}
{\cal G}_Q {\left[ \eta; {\bf P} {\left( \eta \right)} \right]} && = {\frac {2} {\omega_2}} \int \limits_{0}^{\eta} {\rm d} \sigma \sinh \omega_2 {\left( \eta - \sigma \right)} \nonumber 
\\ 
&& \times {\left[ 2 q_0 {\cal P}^2 {\left( \sigma \right)} {\cal Q} {\left( \sigma \right)} + q_1 {\cal Q}^3 {\left( \sigma \right)} \right]}, \label{VecFunctQ}
\end{eqnarray}
where 
\begin{equation}
p_0 = {\frac {\Gamma_{ab}} {{\mathcal a}^2 v_{\omega}^{\prime}}}, \qquad q_0 = {\frac {\Gamma_{ba}} {{\mathcal a}^2 v_{\Omega}^{\prime}}}, \qquad q_1 = {\frac {\Gamma_{bb}} {{\mathcal a}^2 v_{\Omega}^{\prime}}} \label{DubViolConst}
\end{equation}
In addition, ${\cal P}_0$ and ${\cal Q}_0$ are the solutions of the homogeneous parts of Eqs. (\ref{NLinODEP}) and (\ref{NLinODEQ}), respectively.

It can be shown that the solution of the functional equation (\ref{VolterraVec}) can be expressed as \cite{TzenovBOOK} 
\begin{equation}
{\bf P} {\left( \eta \right)} = {\frac {{\widehat{\mathbf{\Gamma}}} \exp {\left\{ {\displaystyle \int \limits_{0}^{\eta}} {\rm d} \lambda {\dfrac {\delta} {\delta {\bf P}_0 {\left( \lambda \right)}}} \cdot {\mathbfcal G} {\left[ \lambda; {\bf P}_0 {\left( \lambda \right)} \right]} \right\}} \bullet {\bf P}_0 {\left( \eta \right)}} {{\widehat{\mathbf{\Gamma}}} \exp {\left\{ {\displaystyle \int \limits_{0}^{\eta}} {\rm d} \lambda {\dfrac {\delta} {\delta {\bf P}_0 {\left( \lambda \right)}}} \cdot {\mathbfcal G} {\left[ \lambda; {\bf P}_0 {\left( \lambda \right)} \right]} \right\}} \bullet {\bf 1}}}. \label{GammaExpSol}
\end{equation}
A few remarks are now in order. Since the solution above is given in a symbolic form, it is necessary to clarify the way it should be applied. The gamma-exponent operator implies that after the exponential function is being developed in a power series (involving multiple integrals) of the vector functional ${\mathbfcal G}$, all functional derivatives should be shifted to the left, so as to act on all functional arguments entering the corresponding expressions. These functional arguments include ${\bf P}_0$s depending on all integration variables, as well as ${\bf P}_0 {\left( \eta \right)}$ figuring out in the numerator. Further details can be found in Refs. \citenum{TzenovBOOK}, \citenum{DubViol1,DubViol2,DubViol3,Arcidosso}. 

\begin{figure}
\begin{center} 
\includegraphics[width=8.0cm]{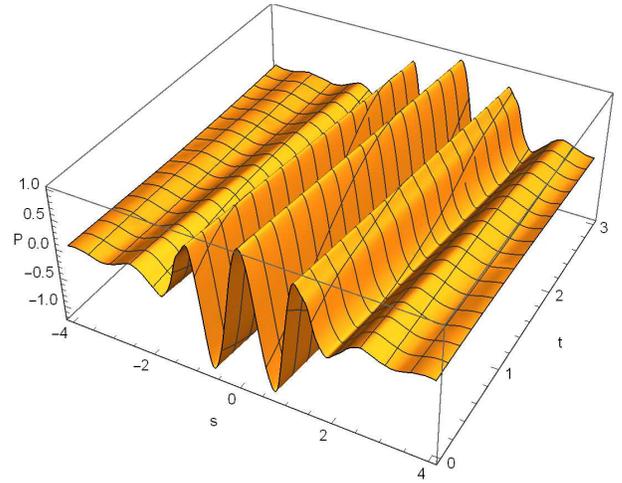}
\caption{\label{fig1:epsart} Evolution of the traveling wave amplitude ${\cal P}$ for the case $k = 1.543613$, $v_T^2 = 0.1$, $\mu = -1.0$, $G = 1.0$ and $H = v_T$.}
\end{center}
\end{figure}

Expanding the gamma-exponent in the numerator, as well as in the denominator of Eq. (\ref{GammaExpSol}) in a formal Volterra series, we represent the latter as a ratio of two series 
\begin{equation}
{\bf P} {\left( \eta \right)} = {\frac {{\displaystyle \sum \limits_{n=0}^{\infty} {\bf P}^{(n)} {\left( \eta \right)}}} {{\displaystyle \sum \limits_{n=0}^{\infty} D^{(n)} {\left( \eta \right)}}}}. \label{DuboisViolette}
\end{equation}
For the first few terms, we obtain 
\begin{equation}
{\cal P}^{(0)} = {\cal P}_0 \qquad \quad {\cal Q}^{(0)} = {\cal Q}_0, \qquad \quad D^{(0)} = 1, \label{DubViolZerOrd}
\end{equation}
\begin{equation}
{\cal P}^{(1)} = {\frac {2 p_0} {\omega_1}} \int \limits_{0}^{\eta} {\rm d} \sigma \sinh \omega_1 {\left( \eta - \sigma \right)} {\cal Q}_0^2 {\left( \sigma \right)} {\cal P}_0 {\left( \sigma \right)}, \label{DubViolFirOrdP}
\end{equation}
\begin{eqnarray}
{\cal Q}^{(1)} = {\frac {2} {\omega_2}} \int \limits_{0}^{\eta} {\rm d} \sigma && \sinh \omega_2 {\left( \eta - \sigma \right)} \nonumber 
\\ 
&& \times {\left[ 2 q_0 {\cal P}_0^2 {\left( \sigma \right)} {\cal Q}_0 {\left( \sigma \right)} + q_1 {\cal Q}_0^3 {\left( \sigma \right)} \right]}, \label{DubViolFirOrdQ}
\end{eqnarray}
\begin{equation}
D^{(1)} = 0, \qquad \qquad D^{(2)} = D^{(2)}_1 + D^{(2)}_2 + D^{(2)}_3, \label{DubViolSecOrdD}
\end{equation}
\begin{eqnarray}
D^{(2)}_1 = - {\frac {2 p_0^2} {\omega_1^2}} \int \limits_{0}^{\eta} {\rm d} \lambda_1 && \int \limits_{0}^{\eta} {\rm d} \lambda_2 \sinh^2 \omega_1 {\left( \lambda_1 - \lambda_2 \right)} \nonumber 
\\ 
&& \times {\cal Q}_0^2 {\left( \lambda_1 \right)} {\cal Q}_0^2 {\left( \lambda_2 \right)}, \label{DubViolSecOrdD1}
\end{eqnarray}
\begin{eqnarray}
D^{(2)}_2 = - {\frac {32 p_0 q_0} {\omega_1 \omega_2}} && \int \limits_{0}^{\eta} {\rm d} \lambda_1 \int \limits_{0}^{\eta} {\rm d} \lambda_2 \sinh \omega_1 {\left( \lambda_1 - \lambda_2 \right)} \nonumber 
\\ 
&& \times \sinh \omega_2 {\left( \lambda_1 - \lambda_2 \right)} {\cal P}_0 {\left( \lambda_1 \right)} {\cal P}_0 {\left( \lambda_2 \right)} \nonumber 
\\ 
&& \times {\cal Q}_0 {\left( \lambda_1 \right)} {\cal Q}_0 {\left( \lambda_2 \right)}, \label{DubViolSecOrdD2}
\end{eqnarray}
\begin{eqnarray}
 D^{(2)}_3 = && - {\frac {2} {\omega_2^2}} \int \limits_{0}^{\eta} {\rm d} \lambda_1 \int \limits_{0}^{\eta} {\rm d} \lambda_2 \sinh^2 \omega_2 {\left( \lambda_1 - \lambda_2 \right)} \nonumber 
\\ 
&& \times {\left[ 2 q_0 {\cal P}_0^2 {\left( \lambda_1 \right)} + 3 q_1 {\cal Q}_0^2 {\left( \lambda_1 \right)} \right]} \nonumber 
\\ 
&& \times {\left[ 2 q_0 {\cal P}_0^2 {\left( \lambda_2 \right)} + 3 q_1 {\cal Q}_0^2 {\left( \lambda_2 \right)} \right]}, \label{DubViolSecOrdD3}
\end{eqnarray}

\begin{figure}
\begin{center} 
\includegraphics[width=8.0cm]{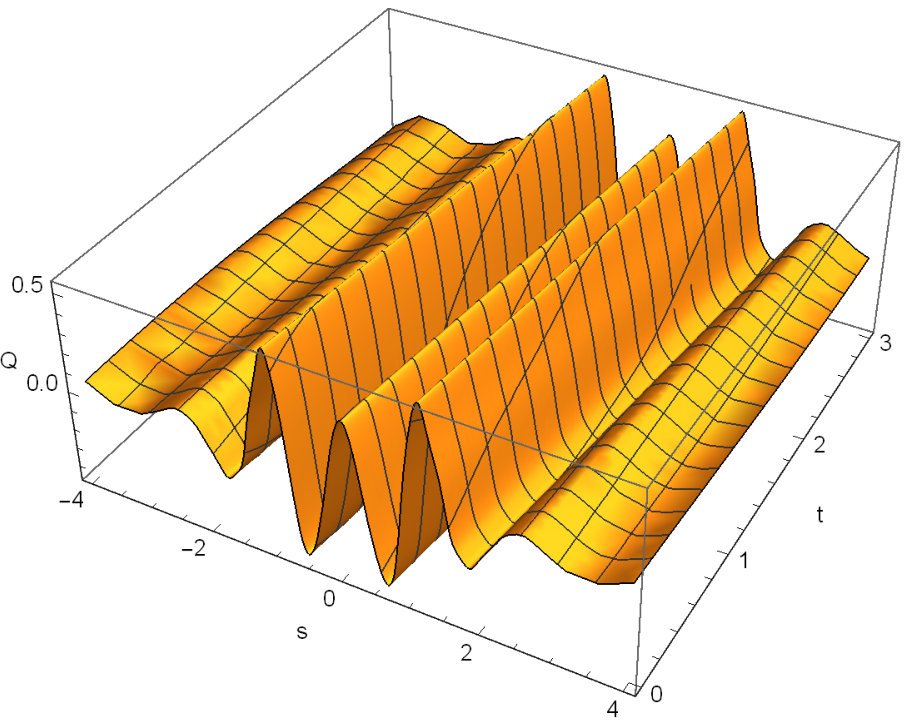}
\caption{\label{fig2:epsart} Evolution of the traveling wave amplitude ${\cal Q}$ for the case $k = 1.543613$, $v_T^2 = 0.1$, $\mu = -1.0$, $G = 1.0$ and $H = v_T$.}
\end{center}
\end{figure}

For negative values of $\mu$, the homogeneous solution ${\bf P}_0 {\left( \eta \right)}$ can be taken in the form 
\begin{equation}
{\cal P}_0 {\left( \eta \right)} = G \cos {\left( \omega_1 \eta + {\mathcal g} \right)}, \quad \quad {\cal Q}_0 = H \cos {\left( \omega_2 \eta + {\mathcal h} \right)}. \label{HomogSolut}
\end{equation}
The evolution of the plasma wave amplitude ${\cal P}$ and the amplitude of the electron current velocity ${\cal Q}$ given by Eq. (\ref{DuboisViolette}) up to second order is shown in Figures \ref{fig1:epsart} and \ref{fig2:epsart}. 

\begin{figure}
\begin{center} 
\includegraphics[width=8.0cm]{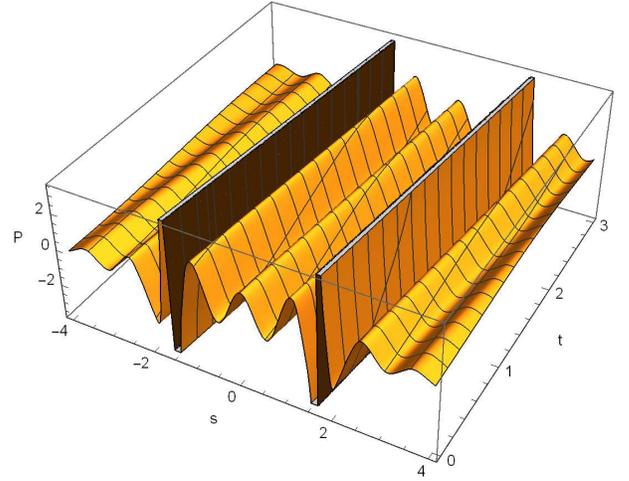}
\caption{\label{fig3:epsart} Evolution of the traveling wave amplitude ${\cal P}$ close to linear resonance $\omega_1 - \omega_2 = 0$. The values of the corresponding parameters are $k = 1.543613$, $v_T^2 = 0.1$, $\mu = -2.0245$, $G = 1.0$ and $H = v_T$.}
\end{center}
\end{figure}

It is important to emphasize that the formal series solution given by Eq. (\ref{DuboisViolette}) is still well behaved even for values of the parameters $k$ and $\mu$ close to linear resonance $\omega_1 - \omega_2 = 0$. The corresponding result is shown in Figure \ref{fig3:epsart}. Up to second order, the numerator in Eq. (\ref{DuboisViolette}) scales as $\eta$, while the denominator scales as $\eta^2$. This holds in both, the resonant and the non-resonant case. Thus, the second order traveling wave solution represents $1 / \eta$-damped quasi-periodic oscillations. 

\section{\label{sec:conclude}Concluding Remarks}

Starting from first principle, an exact relativistic hydrodynamic closure of equations describing the dynamics of various species in a quasi-neutral plasma has been obtained. It has been shown that the set of equations for the macroscopic hydrodynamic variables coupled to the wave equations for the self-consistent electromagnetic field is fully equivalent to the Vlasov-Maxwell system for a special type of relativistic water-bag solutions of the Vlasov equation. As expected, the warm fluid dynamic equations derived here are invariant under Lorentz transformation. Another intriguing feature of our hydrodynamic picture is the underlying pressure law. The latter, combined with the ponderomotive potential (proportional to the transverse vector potential squared) represents an effective relativistic enthalpy of the system. There are not many cases of exact relativistic hydrodynamic closures, and in this sense the results obtained here are of exceptional value.

Based on the method of multiple scales, a further reduction of the macroscopic and the wave equations has been performed. 
This reduction represents by itself a separation of fast (oscillatory) variables from slow ones (called amplitudes or envelopes), which usually govern the formation of stable patterns on longer time scales. Thus, a system comprising a vector nonlinear Schrodinger equation for the transverse envelopes of the self-consistent plasma wakefield coupled to a scalar nonlinear Schrodinger equation for the electron current velocity envelope has been derived. It is noteworthy to mention that our reduction has been performed in the single mode (single value of the wave number $k$) approximation. Generally speaking, it is possible to carry out a full reduction, but the resultant amplitude equations comprise an infinite set of coupled nonlinear Schrodinger equations. In this case, one can think of a gas consisting of mutually interacting quasi-particles. 

Using the method of formal series of Dubois-Violette, a traveling wave solution of the derived set of coupled nonlinear Schrodinger equations in the case of circular wave polarization has been obtained. This solution is represented by a ratio of two formal Volterra series, and is not only compact and elegant but very useful for concrete practical applications. A remarkable property of the formal series solution is the fact that near a resonance the denominator is divergent at least as much as the numerator, so that their ratio gives a reasonable and relevant for applications result. To provide a way of assessing higher order contributions, one needs as many terms in (\ref{DuboisViolette}) as possible. The calculations to obtain the fourth and higher order terms become rather cumbersome, so that computer-aided analytical manipulations are strictly necessary. 

A careful inspection of the results presented in Figures \ref{fig1:epsart}, \ref{fig2:epsart} and \ref{fig3:epsart} shows that the traveling wave solution of the coupled nonlinear Schrodinger equations represents a damping wave (scaling as $1 / \eta$), which on a scale of $3 \div 4 \; c / \omega_e$  can be considered practically completely subdued.

\begin{acknowledgments}
The author wishes to thank Prof. Calin A. Ur and Dr. Guangling Chen for many enlightening discussions and for their encouragement and support. 

The present work has been supported by Extreme Light Infrastructure -- Nuclear Physics (ELI-NP) Phase II, an innovative project co-financed by the Romanian Government and the European Union through the European Regional Development Fund. 
\end{acknowledgments}

\appendix

\section{\label{sec:appsecthird}Equations of Second and Third Order}

The second order equations read as 
\begin{eqnarray}
G_0^2 \partial_t n_2 + && \partial_s v_2 + G_0^2 \partial_{t_1} n_1 + \partial_{s_1} v_1 + {\frac {1} {2}} \partial_t {\left( v_1^2 + A_1^2 \right)} \nonumber 
\\ 
&& + 3 v_T^2 G_0^4 \partial_t n_1^2 + G_0^2 \partial_s {\left( n_1 v_1 \right)} = 0, \label{SecOrdCont}
\end{eqnarray}
\begin{eqnarray}
\partial_t^2 v_2 + && 2 v_T^2 G_0^2 \partial_t \partial_s n_2 + v_2 + 2 \partial_t \partial_{t_1} v_1  \nonumber 
\\ 
&& + 2v_T^2 G_0^2 {\left( \partial_t \partial_{s_1} + \partial_{t_1} \partial_s \right)} n_1   \nonumber 
\\ 
&& + 2 v_T^2 G_0^2 \partial_t^2 {\left( n_1 v_1 \right)} + {\frac {1} {2}} \partial_t \partial_s {\left( v_1^2 + A_1^2 \right)} \nonumber 
\\ 
&& + v_T^2 G_0^4 {\left( 1 + 4 v_T^2 \right)} \partial_t \partial_s n_1^2 = - G_0^2 n_1 v_1, \label{SecOrdMomBal}
\end{eqnarray}
\begin{equation}
{\left( {\boldsymbol{\Box}} - \lambda \right)} {\bf A}_2 + 2 {\left( \partial_s \partial_{s_1} - \partial_t \partial_{t_1} \right)} {\bf A}_1 = {\left( 1 + 2 v_T^2 \right)} n_1 {\bf A}_1. \label{SecOrdWavePerp}
\end{equation}
Using Eq. (\ref{SecOrdCont}), the second order line density $n_2$ can be eliminated from Eq. (\ref{SecOrdMomBal}). As a result, we obtain 
\begin{eqnarray}
&& {\left( \partial_t^2 - 2 v_T^2 \partial_s^2 + 1 \right)} v_2 + 2 {\left( \partial_t \partial_{t_1} v_1 - v_T^2 \partial_s \partial_{s_1} v_1 \right.} \nonumber 
\\ 
&& {\left. + v_T^2 G_0^2 \partial_t \partial_{s_1} n_1 \right)}  + {\frac {1} {2 G_0^2}} \partial_t \partial_s {\left( v_1^2 + A_1^2 \right)} \nonumber 
\\ 
&& + v_T^2 G_0^2 \partial_t \partial_s n_1^2 + G_0^2 {\left( 1 - 2 v_T^2 {\boldsymbol{\Box}} \right)} n_1 v_1 = 0, \label{SecOrdMomBalan}
\end{eqnarray}

The complete third order perturbation equations look very cumbersome. We are not interested here in third order regular solution involving higher harmonics (and third order combinations) of the two basic wave modes. That is why, we shall present below only the fractions of the third order equations, which give rise to secular terms to be cancelled by the multiple scales procedure. These fractions can be written as follows 
\begin{eqnarray}
&& {\left( \partial_{s_1}^2 + 2 \partial_s \partial_{s_2} - \partial_{t_1}^2 - 2 \partial_t \partial_{t_2} \right)} {\bf A}_1 \nonumber 
\\ 
&& = {\left( 1 + 2 v_T^2 \right)} n_1 {\bf A}_2 + {\left[ 2 v_T^2 n_1^2 + {\left( 1 + 2 v_T^2 \right)} n_2  \right]} {\bf A}_1, \quad \label{ThiOrdWavePerp}
\end{eqnarray}
and 
\begin{eqnarray}
&& - 2 v_T^2 G_0^2 \partial_s \partial_{t_1} n_2 - 2 v_T^2 \partial_s {\left( G_0^2 \partial_{t_2} n_1 + \partial_{s_2} v_1 \right)} \nonumber
\\ 
&& + 2 v_T^2 G_0^2 {\left( \partial_t \partial_{s_1} + \partial_s \partial_{t_1} \right)} n_2 + {\left( \partial_{t_1}^2 + 2 \partial_t \partial_{t_2} \right)} v_1 \nonumber 
\\ 
&& + 2 v_T^2 G_0^2 {\left( \partial_t \partial_{s_2} + \partial_{t_1} \partial_{s_1} + \partial_s \partial_{t_2} \right)} n_1 \nonumber
\\ 
&& = - 2 v_T^2 G_0^2 \partial_t \partial_s {\left( n_1 n_2 + 2 v_T^2 G_0^2 n_1^3 \right)} \nonumber
\\ 
&& + 2 v_T^2 G_0^2 {\boldsymbol{\Box}} {\left( n_1 v_2 + n_2 v_1 \right)} + 6 v_T^4 G_0^4 \partial_s^2 {\left( n_1^2 v_1 \right)} \nonumber 
\\ 
&& - v_T^2 {\left( 1 + 4 v_T^2 \right)} G_0^4 \partial_t^2 {\left( n_1^2 v_1 \right)} + v_T^2 \partial_s^2 v_1 {\left( v_1^2 + A_1^2 \right)} \nonumber 
\\ 
&& - {\frac {1} {G_0^2}} \partial_t \partial_s {\left( v_1 v_2 + {\bf A}_1 \cdot {\bf A}_2 \right)} - {\frac {1} {2}} \partial_t^2 v_1 {\left( v_1^2 + A_1^2 \right)} \nonumber 
\\ 
&& - G_0^2 {\left( n_1 v_2 + n_2 v_1 \right)} - 3 v_T^2 G_0^4 n_1^2 v_1 - {\frac {v_1} {2}} {\left( v_1^2 + A_1^2 \right)}. \label{ThiOrdMomBal}
\end{eqnarray}

\section{\label{sec:apphirota}Bilinearization of Eqs. (\ref{NLinSchrodA}) and (\ref{NLinSchrodB}) Using the Non-Conventional Hirota's Direct Method}

The solution of Eqs. (\ref{NLinSchrodA}) and (\ref{NLinSchrodB}) can be obtained by applying Hirota’s bilinearization method, which is a powerful tool for explicit handling of nonlinear partial differential equations. Let $f {\left( z \right)}$ and $g {\left( z \right)}$ be generic functions of the argument indicated. Hirota's bilinear operator is defined according to the relation 
\begin{eqnarray}
{\widehat{D}}_z f {\boldsymbol{\cdot}} g && = {\left. {\left( \partial_z - \partial_{\zeta} \right)} f {\left( z \right)} g {\left( \zeta \right)} \right|}_{\zeta=z} \nonumber 
\\ 
&& = {\left. \partial_{\zeta} f {\left( z + \zeta \right)} g {\left( z - \zeta \right)} \right|}_{\zeta=0} \nonumber 
\\ 
&& = {\left. \partial_{\zeta} {\left[ {\rm e}^{\zeta \partial_z} f {\left( z \right)} \right]} {\left[ {\rm e}^{- \zeta \partial_z} g {\left( z \right)} \right]} \right|}_{\zeta=0} \nonumber 
\\ 
&& = g \partial_z f - f \partial_z g. \label{BilinHirota}
\end{eqnarray}
Generalizations for powers of the Hirota operator, as well as its action on functions of more than one variable is straightforward. 
The correct bilinear equations can be obtained by following a non-conventional method, whose key point is the introduction of an auxiliary function, whose determination is not unique. By virtue of the bilinearizing transformation
\begin{equation}
{\mathbfcal A} = {\frac {\bf G} {F}}, \qquad \qquad  {\cal B} = {\frac {H} {F}}, \label{BilinTrans}
\end{equation}
where ${\bf G}$ and $H$ are complex valued vector and scalar functions, respectively, and $F$ is a real function, the transformed Eqs. (\ref{NLinSchrodA}) and (\ref{NLinSchrodB}) acquire the form
\begin{eqnarray}
&& F {\left[ i {\widehat{D}}_{\eta} + {\frac {{\mathcal a}^2 v_{\omega}^{\prime}} {2}} {\left( {\widehat{D}}_{\xi} - {\widehat{D}}_{\eta} \right)}^2 \right]} {\bf G} {\boldsymbol{\cdot}} F \nonumber
\\ 
&& = {\frac {{\mathcal a}^2 v_{\omega}^{\prime}} {2}} {\bf G} {\left( {\widehat{D}}_{\xi} - {\widehat{D}}_{\eta} \right)}^2 F {\boldsymbol{\cdot}} F \nonumber 
\\ 
&& + \Gamma_{aa} {\bf G}^2 {\bf G}^{\ast} + \Gamma_{ab} {\left| H  \right|}^2 {\bf G}, \label{BasEqTransA}
\end{eqnarray}
\begin{eqnarray}
&& F {\left[ i {\widehat{D}}_{\xi} + {\frac {{\mathcal a}^2 v_{\Omega}^{\prime}} {2}} {\left( {\widehat{D}}_{\xi} - {\widehat{D}}_{\eta} \right)}^2 \right]} H {\boldsymbol{\cdot}} F  \nonumber
\\ 
&& = {\frac {{\mathcal a}^2 v_{\Omega}^{\prime}} {2}} H {\left( {\widehat{D}}_{\xi} - {\widehat{D}}_{\eta} \right)}^2 F {\boldsymbol{\cdot}} F \nonumber 
\\ 
&& + \Gamma_{ba} {\left| {\bf G} \right|}^2 H + \Gamma_{bb} {\left| H \right|}^2 H. \label{BasEqTransB}
\end{eqnarray}
Next, we require that 
\begin{equation}
{\left[ i {\widehat{D}}_{\eta} + {\frac {{\mathcal a}^2 v_{\omega}^{\prime}} {2}} {\left( {\widehat{D}}_{\xi} - {\widehat{D}}_{\eta} \right)}^2 \right]} {\bf G} {\boldsymbol{\cdot}} F = - {\frac {{\cal S} {\bf G}} {F}}, \label{BasicBilin1}
\end{equation}
\begin{equation}
{\left[ i {\widehat{D}}_{\xi} + {\frac {{\mathcal a}^2 v_{\Omega}^{\prime}} {2}} {\left( {\widehat{D}}_{\xi} - {\widehat{D}}_{\eta} \right)}^2 \right]} H {\boldsymbol{\cdot}} F = {\frac {{\cal S} H} {F}}, \label{BasicBilin2}
\end{equation}
\begin{equation}
{\left( {\widehat{D}}_{\xi} - {\widehat{D}}_{\eta} \right)}^2 F {\boldsymbol{\cdot}} F = \Sigma, \label{BasicBilin3}
\end{equation}
where ${\cal S}$ and $\Sigma$ are yet unknown functions to be determined in a while. 

Substitution of Eqs. (\ref{BasicBilin1}) -- (\ref{BasicBilin3}) into Eqs. (\ref{BasEqTransA}) and (\ref{BasEqTransB}) yields a linear system of equations for the two unknowns ${\cal S}$ and $\Sigma$. Solving this system, we obtain 
\begin{eqnarray}
{\cal S} = {\frac {1} {\mathcal b}} && {\left[ {\left( v_{\omega}^{\prime} \Gamma_{bb} - v_{\Omega}^{\prime} \Gamma_{ab} \right)} {\left| H \right|}^2 \right.} \nonumber 
\\ 
&& {\left. + v_{\omega}^{\prime} \Gamma_{ba} {\left| {\bf G} \right|}^2 - {\frac {v_{\Omega}^{\prime} \Gamma_{aa} {\bf G}^2 {\bf G}^{\ast {\bf 2}} } {{\left| {\bf G} \right|}^2}} \right]}, \label{SFunction}
\end{eqnarray}
\begin{equation}
\Sigma = - {\frac {2} {{\mathcal a}^2 {\mathcal b}}} {\left[ {\frac {\Gamma_{aa} {\bf G}^2 {\bf G}^{\ast {\bf 2}} } {{\left| {\bf G} \right|}^2}} + \Gamma_{ba} {\left| {\bf G} \right|}^2 + {\left( \Gamma_{ab} + \Gamma_{bb} \right)} {\left| H \right|}^2 \right]}, \label{SigmaFunction}
\end{equation}
where 
\begin{equation}
{\mathcal b} = v_{\omega}^{\prime} + v_{\Omega}^{\prime}. \label{BCoefficient}
\end{equation}

The set of equations (\ref{BasicBilin1}) -- (\ref{BasicBilin3}) can be solved by introducing the following power series expansions for ${\bf G}$, $H$, $F$ and ${\cal S}$
\begin{equation}
{\bf G} = \epsilon {\bf G}_1 + \epsilon^3 {\bf G}_3 + \dots, \qquad H = \epsilon H_1 + \epsilon^3 H_3 + \dots, \label{HirotaPert1}
\end{equation}
\begin{equation}
F = 1 + \epsilon^2 F_2 + \epsilon^4 F_4 + \dots, \qquad {\cal S} = \epsilon^2 {\cal S}_2 + \epsilon^4 {\cal S}_4 + \dots, \label{HirotaPert2}
\end{equation}
and an expansion for $\Sigma$, similar to the one for ${\cal S}$ above, containing only even powers of $\epsilon$. In analogy with Section \ref{subsec:multiple}, the quantity $\epsilon$ is again a formal small parameter indicating the order of magnitude of various terms in the series expansions introduced above. 

\nocite{*}
\bibliography{aipsamp}

\end{document}